\documentclass[twocolumn,showpacs,preprintnumbers,amsmath,amssymb]{revtex4}
\usepackage{amsfonts}
\usepackage{mathrsfs}
\usepackage{graphicx}
\usepackage{dcolumn}
\usepackage{bm}

\begin{document}

\title{Quantum Information Approach to Rotating Bose-Einstein Condensate}

\author{Zhao Liu, Hongli Guo,  Shu Chen, and Heng Fan}
\affiliation{%
Institute of Physics, Chinese Academy of Sciences, Beijing 100190,
China
}%
\date{\today}

\begin{abstract}
We investigate the 2D weakly interacting rotating Bose-Einstein
condensate by the tools of quantum information theory. The critical
exponents of the ground state fidelity susceptibility and the
correlation length of the system are obtained for the sudden change
of the ground state when the first vortex is formed. This sudden
change can also be indicated by the ground state entanglement. We
also find the single-particle entanglement can be an indicator of
the angular momentums for some real ground states. The
single-particle entanglement of fractional quantum Hall states such
as Laughlin state and Pfaffian state is also studied.
\end{abstract}

\maketitle


{\label{sec:level1}} \section{Introduction}Since the Bose-Einstein
condensate (BEC) was observed in trapped alkali-metal atoms
\cite{BEC1,BEC2,FD}, the response of those systems to rotation has
attracted considerable attention. Unlike classical systems, the BEC
can only gain angular momentum by forming quantized vortices when it
is stirred. When rotation frequency is small enough, no motion of
the system can be observed, while above some critical frequencies,
vortices are formed. These vortices are signatures of superfluidity
\cite{VoFo}. The formation of the first vortex is perhaps especially
interesting. Before the formation of the first vortex, the ground
state of BEC does not rotate while after the formation the ground
state is a single-vortex state, in which all particles rotate around
their mass center. Therefore a macroscopic symmetry breaking must
happen. However, this sudden change of the ground state has not been
studied quantitatively. Some important physical quantities such as
critical exponents are still unknown.

Besides the sudden change of the ground state, many authors focused
on the the ground state energy and the ansatz ground state wave
function when the system gains a fixed angular momentum through
rotation in the weak interaction limit
\cite{NKWa,BGT,NK,RN,NJ,ADJa,AJ,NNJ,ALF}. Generally speaking, the
theoretical methods they used mainly include Gross-Pitaevskii (GP)
mean field theory and exact diagonalization (ED). In GP mean field
theory, the ground state many body wave function is simply expressed
as a product of $N$ single-particle states, namely a non-entangled
state. However, sometimes the mean field theory cannot give a good
description to the system. As pointed out in Ref.\cite{DNMJ}, when
the first vortex is formed, the ground state of the rotating BEC
will change from a product state to a strongly-correlated entangled
state, leading to the invalidity of the description of the mean
field theory. Furthermore, when the number of vortices $N_{v}$ is
large, the mean field theory predicts the ground state of the system
is a vortex lattice phase \cite{BR}. This prediction is correct only
when the filling fraction $\nu\equiv N/N_{v}\gtrsim\nu_{c}$ with
$\nu_{c}\sim6$ \cite{NNJ}, where $N$ is the particle number. When
$\nu\lesssim\nu_{c}$, the ground state is a strongly correlated
vortex liquid phase. Therefore it is the entanglement between
particles that makes GP mean field theory invalid. So studying
entanglement in rotating BEC will be very important to help us
understand the properties of this system, as what has been done in
other condensed matter systems \cite{Vidal}. However, to our
knowledge, this aspect has not yet been investigated.

In this Article, we use the method of ED in the weakly interacting
regime to remain the entanglement property of rotating BEC.
Several tools of quantum information theory are used to investigate
this system. First we use the ground state fidelity and fidelity
susceptibility, which can precisely locate the critical point of a
possibly unknown quantum phase transition
\cite{Sun,PN,WYS,Chen,HQZ,Gu,SNS}, to study the ground state sudden
change when the first vortex is formed. By using finite-size scaling
analysis for even $N$, we obtain the critical exponents of both
fidelity susceptibility and correlation length. We find the ground
state single-particle entanglement can also indicate this sudden
change of the ground state. Then we use the von Neumann entropy to
calculate the single-particle entanglement of the ground states of
subspaces of fixed $z$-component angular momentum $L_{z}$ of the
system. Interestingly, we find that this single-particle
entanglement can indicate some $L_{z}$s of the real ground states,
namely those stable states. Finally, we study the relation between
single-particle entanglement and $N$ for some special subspace
ground states. We find the single-particle entanglement of both
bosonic Pfaffian state and bosonic Laughlin state diverges
logarithmical with $N$, showing a strongly-correlated
characteristics of vortex liquid phase, while the single-particle
entanglement of single-vortex state, namely the ground state in the
subspace $L_{z}=N\hbar$, decays with $N$.

{\label{sec:level1}} \section{Model}In rotating reference frame, the
Hamiltonian of a 2D rotating $N$-boson system with rotation
frequency $\Omega$ trapped in a harmonic oscillator potential is
$\mathcal {H}=\sum_{i=1}^{N}\mathcal {H}_{0,i}+\mathcal {U}$, where
$\mathcal
{H}_{0,i}=-\frac{\hbar^2}{2m}\nabla_{i}^{2}+\frac{1}{2}m\omega^{2}\textbf{\emph{r}}_{i}^{2}-\Omega
\hat{L}_{z,i}$ is the single-particle Hamiltonian and $\mathcal
{U}\propto\sum_{i<j}^{N}\delta(\emph{\textbf{r}}_{i}-\emph{\textbf{r}}_{j})$
is the interaction energy.
Now we suppose the system has a fixed angular momentum $L_{z}=L\hbar
(L\geq0)$, then its ground state energy when the interaction is
absent is $E_{g}=(L+N)\hbar\omega-L\hbar\Omega$.
The corresponding ground state is $\prod_{i=1}^{N}\varphi_{l_{i}}$
with a constrain $\sum_{i=1}^{N}l_{i}=L$, where
$\varphi_{l}=\frac{1}{\sqrt{\pi l!}}z^{l}e^{-|z|^{2}/2}$ with
$z=x+\textmd{i}y$ is the normalized single-particle lowest Landau
level(LLL) wave function of $\mathcal {H}_{0,i}$. If the interaction
is weak enough (this means in Eq.(\ref{e1})
$NU_{0}\lesssim\hbar\omega$, throughout the calculation we make
$NU_{0}=\hbar\omega/2$ to keep the validity of the LLL
approximation), the dynamics of the system is restricted in the LLL,
from which other Landau levels are separated by a large energy gap
$2\hbar\omega$, so that we can use $\varphi_{l}$ to do the second
quantization of the Hamiltonian, leading to
\begin{eqnarray}
\mathcal {H}_{L}=(L+N)\hbar\omega-L\hbar\Omega
+U_{0}\sum_{i,j,k,l}U_{i,j,k,l}a_{i}^{\dagger}a_{j}^{\dagger}a_{k}a_{l}
\label{e1}
\end{eqnarray}
with
$U_{i,j,k,l}=\frac{1}{2^{i+j}}\frac{(i+j)!}{\sqrt{i!j!k!l!}}\delta_{i+j,k+l}$.
$a_{l}^{\dagger}$ $(a_{l})$ creates (annihilates) a particle in the
state $\varphi_{l}$. The basis $\mathfrak{B}_{L}$ of the Hilbert
subspace of our system with fixed $L_{z}=L\hbar$ in Fock
representation is $|N_{0}N_{1}...N_{L}\rangle$ with the constrains
that $\sum_{l=0}^{L}N_{l}=N$ and $\sum_{l=0}^{L}(lN_{l})=L$. Under
this basis, we can diagonalize the Hamiltonian (\ref{e1}) to find
its subspace ground state $|\Psi_{0,L}\rangle$ and the ground state
energy $E_{0,L}(\Omega)$.

Now let's consider the total Hamiltonian $\mathcal
{H}=\bigoplus_{L}\mathcal {H}_{L}$. When $\Omega$ varies from 0 to
$\omega$, the one among all $|\Psi_{0,L}\rangle$ with the lowest
$E_{0,L}(\Omega)$ is the real ground state, namely the stable state
$|\Psi_{0}\rangle$ of $\mathcal {H}$. The $L_{z}$ of
$|\Psi_{0}\rangle$, denoted by $L_{z,0}$, forms a series of sharp
steps from 0 to $N(N-1)$ (in $\hbar$ unit from now on) \cite{NK,NJ}.
In the first step at $\Omega=0.75\omega$, $L_{z,0}$ varies from 0 to
$N$, corresponding to the formation of the first vortex, where the
ground state changes suddenly from a non-rotating state to a
single-vortex state.

{\label{sec:level1}} \section{Ground state fidelity and fidelity
susceptibility}At the beginning we consider a general Hamiltonian
$\mathcal {H}(\lambda)=\mathcal {H}_{a}+\lambda\mathcal {H}_{b}$,
where $\lambda$ is a parameter that can be changed, then the ground
state fidelity is defined as
$F=|\langle\Psi_{0}(\lambda+\delta\lambda)|\Psi_{0}(\lambda)\rangle|$.
The ground state fidelity susceptibility can be calculated from the
formula \cite{WYS,Chen,Gu}
\begin{eqnarray}
\chi(\lambda)=-\lim_{\delta\lambda\rightarrow0}\frac{2\ln
F}{\delta\lambda^{2}}=\sum_{n\neq0}\frac{|\langle\Psi_{n}(\lambda)|H_{b}|\Psi_{0}(\lambda)\rangle|^{2}}{[E_{n}(\lambda)-E_{0}(\lambda)]^{2}},
\label{e2}
\end{eqnarray}
where $|\Psi_{0}(\lambda)\rangle$ ($|\Psi_{n}(\lambda)\rangle$) is
the ground (excited) state of $\mathcal {H}(\lambda)$ and
$E_{0}(\lambda)$ ($E_{n}(\lambda)$) is the ground (excited) state
energy. It's obvious that in our system, $\lambda=\Omega$ and
$\mathcal {H}_{b}=\hat{L_{z}}$. For the system with Hamiltonian
$\mathcal {H}=\bigoplus_{L}\mathcal {H}_{L}$, an energy level
crossing of the ground state exists at $\Omega=0.75\omega$ therefore
the ground state fidelity shows a simple drop at
$\Omega=0.75\omega$. One should notice that Eq.(\ref{e2}) is valid
only when there is no degeneracy for the ground state of the system.
So to study the ground state fidelity susceptibility, we have to
eliminate this energy level crossing first. We adopt the method used
in Ref.\cite{DNMJ} to add the stirring potential $\mathcal
{V}\propto \sum_{i=1}^{N}(x_{i}^{2}-y_{i}^{2})$ to the Hamiltonian
$\mathcal {H}$ to generate an energy gap around $\Omega=0.75\omega$.
Then we diagonalize $\mathcal {H}$ which has been added by $\mathcal
{V}$ in the basis $\mathfrak{B}=\bigcup_{L=0}^{N+2}\mathfrak{B}_{L}$
to calculate $\chi(\Omega)$. $\mathcal {V}$ must be small enough to
guarantee that we can still use $\varphi_{l}$, the single-particle
LLL wave function of $\mathcal {H}_{0,i}$ in the absence of
$\mathcal {V}$ to do the second quantization. $\mathcal {V}$ can be
expressed as
$V_{0}\sum_{l}\Big(\sqrt{l(l-1)}a_{l}^{\dagger}a_{l-2}+\sqrt{(l+1)(l+2)}a_{l}^{\dagger}a_{l+2}\Big)$,
where $V_{0}\ll\hbar\omega$ ($V_{0}=0.003\hbar\omega$ in this
section). We find that after adding $\mathcal {V}$, the ground state
degeneracy for odd $N$ is not broken completely, meaning the ground
state fidelity still shows a drop. But for even $N$, the ground
state is non-degenerate so the ground state fidelity is a smooth
curve (FIG.1 and FIG.2). We focus on even $N$ in the following.
\begin{figure}
\includegraphics[height=3cm,width=\linewidth]{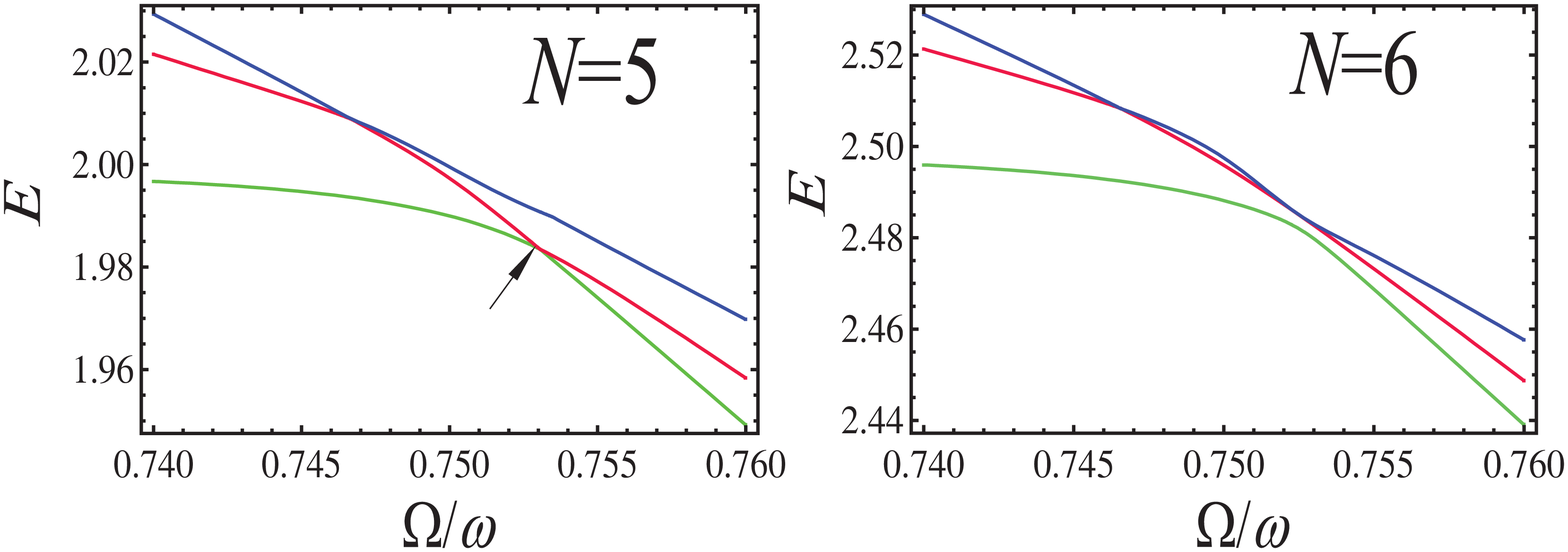}
\caption{\label{fig:epsart} (Color online) The lowest three energy
levels of $\mathcal {H}$ after added by $\mathcal {V}$ for $N=5$ and
$N=6$ (in $\hbar\omega$ unit). The arrow points to the ground state
degeneracy when $N=5$.}
\end{figure}

\begin{figure}
\includegraphics[height=3cm,width=\linewidth]{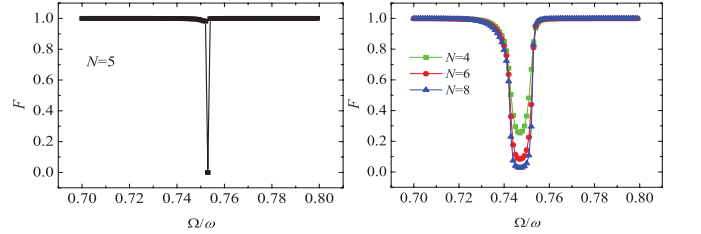}
\caption{\label{fig:epsart} (Color online) The ground state fidelity
$F$ for $N=5 (\delta\Omega=0.001\omega)$ and $N=4,6,8
(\delta\Omega=0.01\omega)$.}
\end{figure}

We calculate $\chi(\Omega)$ for even $N$. We find that
$\chi(\Omega)$ shows a singularity at $\Omega\approx0.7525\omega$,
where a sudden change of the ground state is indicated (FIG.3(a)).
\begin{figure}
\includegraphics[height=6cm,width=1\linewidth]{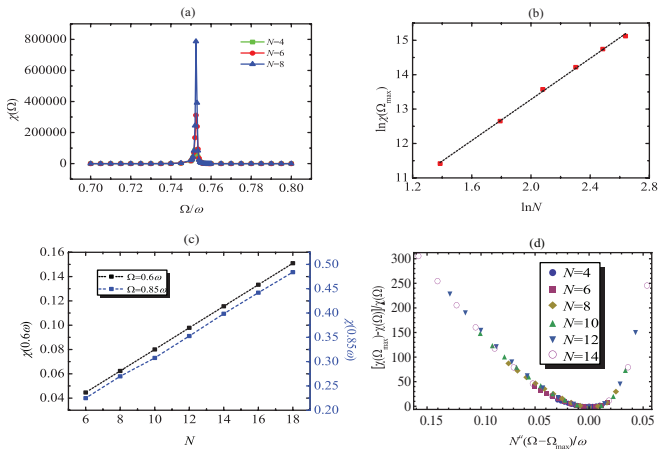}
\caption{\label{fig:epsart} (Color online) The finite-size scaling
analysis of $\chi(\Omega)$ for even $N$.
}
\end{figure}
When studying the sudden change of the ground state, it's necessary
to consider the problem of scaling. First we will study how does the
fidelity susceptibility scale with $N$ when $\Omega$ is at and far
from $\Omega_{\max}$ where the peak locates. The height of the peak
of $\chi(\Omega)$ at $\Omega=\Omega_{\max}$ diverges with $N$ and
scales like $\chi(\Omega_{\max})\propto N^{d_{c}}$ with
$d_{c}\approx2.9862$ (FIG.3(b)). When $\Omega$ is far from (either
below or above) $\Omega_{\max}$, $\chi(\Omega)\propto N^{d}$ with
$d=1$ (FIG.3(c)). In the thermodynamic limit,
$\Omega_{\max}\rightarrow\Omega_{c}$. We are interested in the
critical exponent $\mu$ of the correlation length when $\Omega$
approaches $\Omega_{c}$. It's known that the rescaled fidelity
susceptibility
$\frac{\chi(\Omega_{\max})-\chi(\Omega)}{\chi(\Omega)}$ is a
universal function of $N^{\mu}(\Omega-\Omega_{\max})$ \cite{Gu}.
FIG.3(d) shows this function for $N=4,6,8,10,12,14$ with
$\mu\approx1.35$. We can see that all points of different $N$ locate
on a single curve. At last we also want to know the critical
exponent of $\chi(\Omega)$ when $\Omega$ approaches $\Omega_{c}$.
Near $\Omega_{c}$, $\chi(\Omega)$ must behave like
$\chi(\Omega)\propto1/|\Omega-\Omega_{c}|^{\alpha}$, where
$\alpha=(d_{c}-d)/\mu$ \cite{Gu}. So we can obtain
$\alpha\approx(2.9862-1)/1.35\approx1.4713$. We also find the ground
state single-particle entanglement $S_{1}$, which will be defined
and discussed in detail in the next section, and its first order
derivative $\frac{dS_{1}}{d(\Omega/\omega)}$, which shows a
singularity at $\Omega\approx0.7525\omega$, can indicate the sudden
change of the ground state as well (FIG.4).

\begin{figure}
\includegraphics[height=3cm,width=\linewidth]{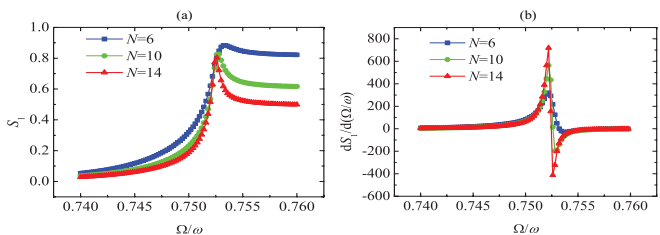}
\caption{\label{fig:epsart} (Color online) The single-particle
entanglement $S_{1}$ of the ground state and its derivative
$\frac{dS_{1}}{d(\Omega/\omega)}$ for $N=6,10,14$.
}
\end{figure}

{\label{sec:level1}} \section{Single-particle entanglement}In this
section, we want to study the ground state single-particle
entanglement when the system has a fixed angular momentum $L_{z}$
and the stirring potential $\mathcal {V}$ is canceled. At the
beginning, we first introduce the definition of entanglement between
particles. For a general $N$-particle pure state
$|\Phi\rangle_{1,2,...,N}$, the entanglement between $n$ particles
and the other $N-n$ particles can be defined as the von Neumann
entropy of the reduced density matrix of the $n$ particles, namely
$S_{n}=-\textrm{Tr}(\rho_{n}\ln\rho_{n})$ with
$\rho_{n}=\textrm{Tr}_{n+1,n+2,...,N}(|\Phi\rangle_{1,2,...,N}\langle\Phi|)$,
where we have supposed that $|\Phi\rangle_{1,2,...,N}$ is invariant
under permutation of particles (this is satisfied in this paper
because of the symmetry of boson wave function) so that $\rho_{n}$
is irrelevant with the choice of the $N-n$ particles in the partial
trace. Therefore to calculate the ground state single-particle
entanglement when the system has a fixed angular momentum $L_{z}$,
we need to digonalize $\mathcal {H}_{L_{z}}$ numerically under the
basis $\mathfrak{B}_{L_{z}}$ to solve the subspace ground state
$|\Psi_{0,L_{z}}\rangle$. In fact $|\Psi_{0,L_{z}}\rangle$ is
expressed in angular momentum occupation Fock representation, which
should be transformed to single-particle state representation. For
example, we have two identical bosons, one of which has angular
momentum 0 and the other has angular momentum 1. In angular momentum
occupation Fock representation, this state is $|11\rangle$, but in
single-particle state representation, this state is
$\frac{1}{\sqrt{2}}(|\varphi_{0}\rangle_{1}|\varphi_{1}\rangle_{2}+|\varphi_{1}\rangle_{1}|\varphi_{0}\rangle_{2})$,
 where $|\varphi_{0}\rangle_{1}|\varphi_{1}\rangle_{2}$ means the first particle is in the state $\varphi_{0}$
 and the second particle is in the state $\varphi_{1}$ ($\varphi_{l}=\frac{1}{\sqrt{\pi
 l!}}z^{l}e^{-|z|^{2}/2}$). After completing this transformation, we trace out $N-1$ particles (from the second to $N$th for simplicity). Finally, the single-particle reduced
density operator can be expressed in the form
$\rho_{1}=\sum_{i=0}^{L_{z}}\xi_{i}|\varphi_{i}\rangle\langle\varphi_{i}|$
and $S_{1}=-\sum_{i=0}^{L_{z}}\xi_{i}\ln\xi_{i}$. Once $L_{z}$ is
fixed, according to Eq.(\ref{e1}) the subspace ground state
$|\Psi_{0,L_{z}}\rangle$ is uniquely determined by the operator
$\sum_{i,j,k,l}U_{i,j,k,l}a_{i}^{\dagger}a_{j}^{\dagger}a_{k}a_{l}$.
Therefore $S_{1}$ is irrelevant with $U_{0}$.

We calculate the ground state single-particle entanglement $S_{1}$
with fixed $L_{z}=1,2,...,N(N-1)+1$. We find that $S_{1}$ has a
tendency to grow with $L_{z}$ but the whole curve shows a behavior
of slight oscillation. To see this oscillation clearer, in the
following we subtract a second order polynomial in $L_{z}$ from
$S_{1}(L_{z})$. For example, for $N=8$ we subtract
$-0.0008L_{z}^{2}+0.0817L_{z}+0.5057$ from $S_{1}(L_{z})$. In FIG.5,
we find there exist a series of local minima at $L_{z}=L_{m}$ for
$S_{1}(L_{z})$. Interestingly, most of these $L_{m}$ are either the
real ground state angular momentum $L_{z,0}$ or the candidate of
$L_{z,0}$ predicted by the composite fermion theory \cite{NK}.
Therefore the calculation of $S_{1}$ of the subspace ground state
$|\Psi_{0,L_{z}}\rangle$ may give a way of electing some $L_{z}$s of
real ground state $|\Psi_{0}\rangle$ (TABLE I).

\begin{table*}
\caption{\label{tab:table1} In this TABLE, we list $L_{m}$, where
$S_{1}(L_{z})$ shows a local minimum; $L_{z,0}$, the real ground
state angular momentum, and $L_{\textrm{CF}}$, the candidate of
$L_{z,0}$ predicted by the composite fermion theory \cite{NK} for
$N=5,6,7,8$.}
\begin{ruledtabular}
\begin{tabular}{cccccccc}
$N$&$L_{m}$&$L_{z,0}$&$L_{\textrm{CF}}$\\
\hline
5&5,8,10,12,15,20&5,8,10,12,15,20&5,8,10,12,15,20\\
6&6,10,12,15,18,20,24,30&6,10,12,15,20,24,30&6,10,12,15,18,20,24,30\\
7&7,12,14,18,20,22,24,27,30,35,42&7,12,15,18,24,30,35,42&7,12,15,18,20,22,24,27,30,35,42\\
8&8,14,16,18,21,24,28,30,32,35,38,42,48,56&8,12,14,18,24,30,32,35,42,56&8,14,18,21,24,26,28,30,32,35,38,42,48,56\\
\end{tabular}
\end{ruledtabular}
\end{table*}

\begin{figure}
\includegraphics[height=7cm,width=\linewidth]{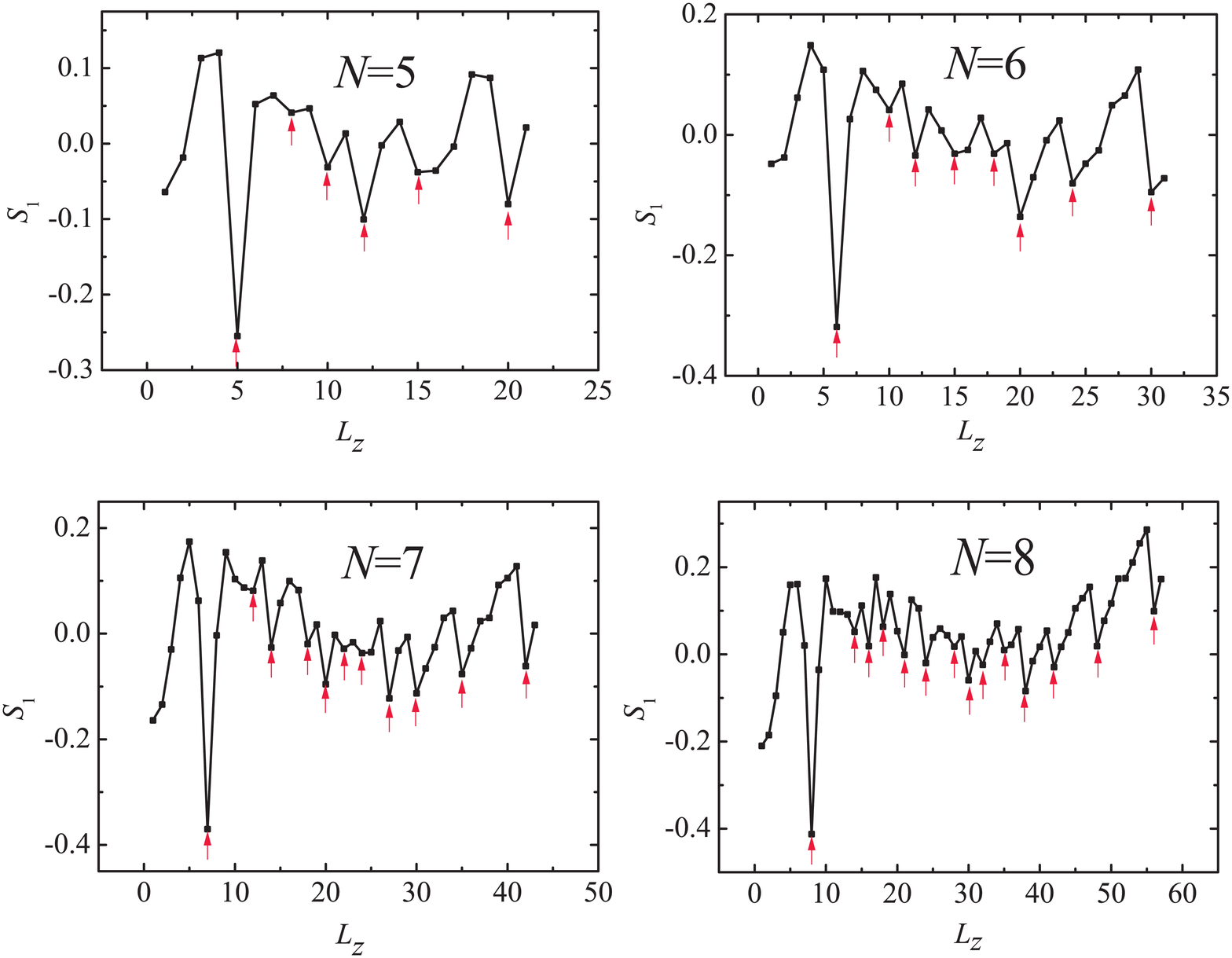}
\caption{\label{fig:epsart} (Color online) The single-particle
entanglement $S_{1}$ of the ground state of the subspaces of fixed
$L_{z}$. A second order polynomial of $L_{z}$ has been subtracted
from $S_{1}(L_{z})$ to make the oscillation clearer. The red arrows
point to the positions of local minima.}
\end{figure}

Now we focus on the $S_{1}$ of three subspace ground states with
special $L_{z}$s. The first one is $L_{z}=N$, which belongs to
$L_{z,0}$ for any $N$. When $L_{z}=N$, the ground state wave
function was conjectured as
$\psi_{N}=\prod_{i=1}^{N}[(z_{i}-z_{c})e^{-|z_{i}|^{2}/2}]$, where
$z_{c}=\frac{1}{N}\sum_{i=1}^{N}z_{i}$ is the position of the center
of mass. The single-particle reduced density matrix to accuracy
$\mathcal {O}(1/N^{2})$ is \cite{NKWa}
\begin{eqnarray}
\rho_{1}(z,z')=\int\Big(\prod_{i=2}^{N}dz_{i}\Big)\psi_{N}(z,z_{2},...,z_{N})
\psi_{N}^{*}(z',z_{2},...,z_{N})\nonumber\\
=\frac{1}{N}\varphi_{0}(z)\varphi_{0}^{*}(z')+\Big(1-\frac{2}{N}\Big)\varphi_{1}(z)\varphi_{1}^{*}(z')
+\frac{1}{N}\varphi_{2}(z)\varphi_{2}^{*}(z').\nonumber
\end{eqnarray}
The single-particle entanglement can be obtained as
$S_{1}^{L_{z}=N}\approx-2/N\ln(1/N)-(1-2/N)\ln(1-2/N)$. FIG.6(a)
shows the $S_{1}^{L_{z}=N}$ decays with the growth of $N$. This is
because when $N\rightarrow\infty$, all particles condensate to the
single-particle state $\varphi_{1}$. We know that perfect condensate
has zero particle entanglement \cite{Haque1}.

Next we study other two cases which have close relation with quantum
Hall effect \cite{NTH,SV,AGD}. One is when $L_{z}=\frac{1}{2}N(N-2)$
for even $N$ and $L_{z}=\frac{1}{2}(N-1)^{2}$ for odd $N$. This
subspace ground state has a high overlap with the bosonic Pfaffian
state of $\nu=1$ \cite{NJ}:
$\psi_{\texttt{Pf}}=\prod_{i<j}^{N}\Big[(z_{i}-z_{j})\textrm{Pf}\Big(\frac{1}{z_{i}-z_{j}}\Big)\Big]\prod_{i=1}^{N}e^{-|z_{i}|^{2}/2}$.
Our numerical result is $S_{1}\approx\ln(1.1037N-1.1369)$. The other
is when $L_{z}=N(N-1)$, which belongs to $L_{z,0}$ for any $N$. This
subspace ground state is bosonic Laughlin state of $\nu=1/2$:
$\psi_{\textrm{Lau}}=\prod_{i<j}^{N}(z_{i}-z_{j})^2\prod_{i=1}^{N}e^{-|z_{i}|^{2}/2}$.
Our numerical result is $S_{1}\approx\ln(1.941N-1.081)$. We can find
the single-particle entanglement analytically for some quantum Hall
effect states in spherical geometry obtaining
$S_{1}^{\textrm{Lau}}=\ln(2N-1)$ for $\nu=1/2$ and
$S_{1}^{\textrm{Pf}}=\ln(N-1)$ for $\nu=1$ \cite{Haque2}. Now we are
dealing with a system in disk geometry so our results are not
exactly the same with this. However, the difference is not very
large (FIG.6(b)).

\begin{figure}
\includegraphics[height=4cm,width=\linewidth]{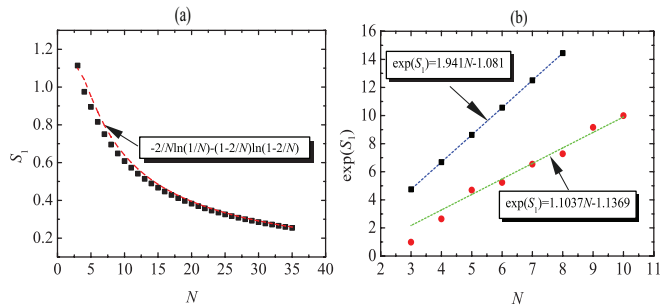}
\caption{\label{fig:epsart} (Color online) (a) $S_{1}$ of the ground
state of the subspace of $L_{z}=N$.
(b) red circle: $S_{1}$ of the ground state of the subspace of
$L_{z}=\frac{1}{2}N(N-2)$ for even $N$ and
$L_{z}=\frac{1}{2}(N-1)^{2}$ for odd $N$, which has a high overlap
with bosonic Pfaffian state.
black cubic: $S_{1}$ of the ground state of the subspace of
$L_{z}=N(N-1)$, which is bosonic Laughlin state.
}
\end{figure}

It's known that when $N$ and $N_{v}$ are both large and our system
has a fixed $L_{z}$, if $\nu=N/N_{v}=N^{2}/(2L_{z})\gtrsim\nu_{c}$
with $\nu_{c}\sim6$, the ground state is a vortex lattice state,
otherwise the ground state is a vortex liquid state. In the vortex
liquid regime for $\nu=1/2$ and $\nu=1$, the results above show that
in the thermodynamic limit $S_{1}$ of the ground state is
logarithmical divergent with $N$. But what's the relation between
$S_{1}$ of the ground state and $N$ for some $\nu$ in the vortex
lattice regime? Considering in this regime the mean field theory
describes our system well, we conjecture that when
$N\rightarrow\infty$, $S_{1}$ will not diverge with $N$ for some
$\nu$ in the vortex lattice regime.

We hope the ground state sudden change when the first vortex is
formed and the entanglement in it can be investigated by
experiments. To achieve this goal, an energy gap between the ground
state and excited states needs to be generated by stirring potential
(see FIG.1). Considering this energy gap decays with the particle
number, the number of particles in experiments should be restricted.
On the other hand, for most experiments of rotating BEC realized in
the laboratory, the filling factor $\nu\sim10^{3}$, well inside the
vortex lattice phase. Experimentalists have elaborate techniques to
observe vortex lattice, while how to detect the vortex liquid state
is still a challenge. However, some experimental methods have been
proposed \cite{Bloch}. We hope the strongly-correlated
characteristics of bosonic Laughlin and Pfaffian state as we show
here by their logarithmical divergent single-particle entanglement
can be verified in experiments in the future.

{\label{sec:level1}} \section{Summary}In summary, we investigate a
2D rotating BEC by tools of quantum information theory. The critical
exponents of ground state fidelity susceptibility and the
correlation length are obtained for the ground state sudden change
when the first vortex is formed. We find the single-particle
entanglement $S_{1}$ of the ground state can be used to detect this
sudden change. We also find a novel property that $S_{1}$ can
indicate some angular momentum $L_{z}$ of the real ground states,
namely those stable states. At last, $S_{1}$ of the ground states in
some special subspaces of fixed $L_{z}$ are calculated to show the
strongly-correlated property of vortex liquid phase. Thus some basic
properties underlying in rotating BEC are clarified and viewed from
the point of quantum information. On the other hand, the formation
of the first vortex in rotating BEC may provide a signature for the
macroscopic entanglement \cite{vedral}.

This work is partially supported by 973 grant No. 2010CB922904.

\end{document}